\documentclass[aps,pre,twocolumn,10pt,showpacs]{revtex4-1}
\usepackage[cp 1252]{inputenc}
\usepackage{amsmath}
\usepackage{graphicx}
\usepackage{srcltx}
\newcommand{\vermelho}{}
\usepackage{fancyhdr}
\begin{document}

\title{Physiological Aging as an Infinitesimally Ratcheted Random Walk}
\author{Bernardo A. Mello}
\email{bernardo@fis.unb.br}
\affiliation{Institute of Physics, University of
Brasilia, 70919-970, Brasilia, DF, Brazil}










\begin{abstract}
The distribution of a population throughout the physiological age of the
individuals is very relevant information in population studies. It has been
modeled by the Langevin and the Fokker-Planck equations. A major problem with
these equations is that they allow the physiological age to move back in time.
This paper proposes an Infinitesimally ratcheted random walk as a way to solve
that problem. Two mathematical representations are proposed. One of them uses a
non-local scalar field. The other one is local, but involves a multi-component
field of speed states. These two formulations are compared to each other and to
the Fokker-Planck equation. The relevant properties are discussed. The dynamics
of the mean and variance of the population age resulting from the two proposed
formulations are obtained.\\

Published as: PHYSICAL REVIEW E 82, 021918 (2010)

DOI: 10.1103/PhysRevE.82.021918

\end{abstract}

\maketitle

\section{Introduction}

The mathematical study of populations started with Malthus \cite{Mal}, who, in
1789, declared
that the human population grows exponentially. Verhulst realized, in 1838, that
the limited
availability of resources prevents that growth and proposed the now famous
logistic equation \cite{Ver}. Further progress was made with the predator-prey
equations, independently proposed by Lotka and Volterra \cite{Lot,Vol} in 1925
and 1926.

In these formulations, each species is described by one variable that
represents the total number of individuals. It is often necessary to have
information about the age structure of the population. In 1945 Leslie used a
matrix containing the probability of surviving or reproducing until the next
time step \cite{Les}. When that matrix is applied to a vector describing the
number of individuals at each discrete age, the vector with the population at
the next time step is obtained.

McKendrick \cite{Ken}, in 1926, and Von Forest \cite{Foe}, in 1959, expressed
age by using a continuous variable. The age structure at time $t$ is replaced
by the probability density  $\hat{p}(a,t)$ where $a$ is the chronological age
(c-age), i.e., the time elapsed from the birth of the individuals and the
instant $t$. The dynamics of a population with a mortality rate
 $\hat{\kappa}(a)$ is described by the convective equation
\begin{equation} \label{Foe}
  \frac{\partial \hat{p}(a,t)}{\partial t} =
	-\frac{\partial \hat{p}(a,t)}{\partial a} - \hat{\kappa}(a)
\hat{p}(a,t).
\end{equation}

Determining the c-age is not possible in several practical situations. In these
cases, the population is divided into development classes depending on some
characteristics of the individuals. In spite of the fact that these classes are
related to the c-age, they are not an exact measurement of it. Since the
sojourn time within a class changes from one individual to another, Lefkovitch,
in 1965, described survival and progression from one class to the next by a
transition matrix, in a Markovian representation of aging \cite{Lef}.

The interaction of the individuals with the environment, among themselves and
with other species depends on the physiological aspects measured by the
physiological age (p-age, $\phi$), through which they are related to the c-age. For that
reason VanSickle, in 1977, proposed using the p-age to describe the population
\cite{Sic}. The p-age $\phi$ can be represented, for example, by the
individual's weight or size, whose average increases monotonically with the c-age $a$. With the
relation between these variables  described by $\phi=\Phi(a)$, VanSickle rewrote
Eq. (\ref{Foe}) as
\begin{equation} \label{vansickle}
  \frac{\partial p(\phi,t)}{\partial t} =
	-\frac{ \partial v(\phi)p(\phi,t)}{\partial \phi} - \kappa(\phi)
p(\phi,t),
\end{equation}
where
\begin{align}
v(\phi) &= \left.\frac{d \Phi}{da}\right|_{\Phi^{-1}(\phi)},\\
p(\phi,t) &= \frac{\hat{p}(\Phi^{-1}(\phi),t)}{v(\phi)},
\text{ and}\\
\kappa(\phi) &= \hat{\kappa}(\Phi^{-1}(\phi)) .
\end{align}

Two important measures of the population distribution are the mean value of
$\phi$,
\begin{equation} \label{mu}
  \mu(t) \equiv \int \phi \, p(\phi,t)\, d\phi\\,
\end{equation}
and the variance of $\phi$
\begin{equation} \label{sig2}
\sigma^2(t) \equiv \int (\phi-\mu)^2\, p(\phi,t)\, d\phi ,
\end{equation}
with $p(\phi,t)$ normalized. None of the above approaches has a parameter that
allows to fit the model to the experimentally measured variance.

The variability of the development rate among individuals was introduced by Lee
et al in 1976  \cite{LeeAl}. After that, much work was done with the
individual-based model of Huston, deAngelis and Post \cite{HusAngPos}.
Kirkpatrick \cite{Kir}, Clother and Brindley \cite{CloBri} exploited the Ito
equation with a Gaussian noise with mean zero, best known among physicists as
the Langevin equation. In \cite{PlaWil}, Plant and Wilson studied populations
with continuous dynamics within stages and discontinuous stage structures. A
formulation of this model using partial derivatives, the Fokker-Planck
equation, was utilized by Buffoni and Pasquali \cite{BufPas}
\begin{equation} \label{FokkerPlanck}
 \frac{\partial p(\phi,t)}{\partial t} =
	-v \frac{\partial p(\phi,t)}{\partial \phi}
	+ \frac{D}{2} \frac{\partial^2 p(\phi,t)}{\partial \phi^2}.
\end{equation}

The time evolution of the p-age depends on several environment factors. The
metabolic rates of insects, for example, change with the temperature. For that
reason the variables  $v$ and $D$ in Eq. (\ref{FokkerPlanck}) could be
functions of the temperature, $T$. Despite this fact, they will be kept
constant from now on.

{\vermelho A method that has being applied in practical situations is the
calculus of degree-days \cite{RolAl} between the occurrence of a biofix (a
detectable biological event) at time $t_0$ and the development stage at time
$t_1$. Formally it is given by
\begin{equation}
DD \equiv \int_{t_0}^{t_1} (T(t)-T_0) H(T(t)-T_0) \, dt ,
\end{equation}
where $H(x)$ is the Heaviside step function, $T$ is the environment
temperature and $T_0$ is the baseline temperature.
Since the development rate is believed to be approximately
proportional to $T-T_0$, the number of degree-days is proportional to the
physiological development. The peak of an infestation is used as the biofix and
the evolution of the population surge is evaluated by the integral. It is,
notwithstanding, not a population description, but just an estimate of the
development state of a population sample.}

The p-age can also be an abstract indicator of the individuals' maturity
\cite{BufPas}. That description is particularly suitable for species with
well-defined stages \cite{NisGur}. Although that abstract value cannot be
determined for every individual, specific values of the p-age are related to
unmistakable biological events, such as eggs hatching, emergence from pupae,
etc. Since the Fountain of Youth has not yet been discovered, it makes no sense
to allow p-age to move backward. Even if such backward movements could happen
at molecular level, they are forbidden at the thermodynamic limit represented
by the p-age of an individual. Therefore, the Langevin and the Fokker-Planck
equations are inappropriate descriptions.

A more suitable mathematical model would be a continuous variable that
moves forward with steps of random, positive length. The physical model for such
a system is an infinitesimally ratcheted random walk (IRRW). It is not a
Brownian ratchet in the sense introduced by Feynman \cite{Fey}, because, in the
present approach: $a$) backward steps are completely forbidden, not just less
probable; $b$) the ratchet teeth are infinitesimal, meaning that even the
smallest movement advances the ratchet to another tooth; $c$) no reference is
made to mechanics or the thermodynamics theory.

Two mathematical models for the IRRW are presented below, disregarding the
boundary effects, reproduction, and death, which results in probability
conservation. As will be shown, in both models the values of  $\mu$ and
$\sigma^2$ evolve uniformly on time:
\begin{equation} \label{dmdt}
 \frac{d\mu(t)}{dt} = v \hspace{1cm}
 \frac{d\sigma^2(t)}{dt} = D .
\end{equation}
In each of the models proposed, expressions for $v$ and $D$ are found.

\section{Non-local formulation}

The simplest approach to the IRRW involves the scalar field $p(\phi,t)$. Since
we are not concerned with inter-particle interactions, the equation must
include linear terms only. Furthermore, the dynamics must be translationally
invariant on $\phi$, at least inside a given development stage. The most
general equation satisfying these requirements is
\begin{equation} \label{intEq}
 \frac{\partial p(\phi,t)}{\partial t} = - \alpha p(\phi,t)
	+ \alpha \int p(\phi',t) f(\phi-\phi') d\phi'.
\end{equation}
Conservation imposes
\begin{equation}
\int f(\phi) d\phi = 1 .
\end{equation}

The Fokker-Planck equation corresponds to having $\alpha=1$ and
\begin{equation}
f(\phi) = \delta(\phi) + v d \frac{\delta(\phi)}{d\phi} -
	D \frac{d^2\delta(\phi)}{d\phi^2} .
\end{equation}
The space-discretized form of Eq. (\ref{intEq}) is
\begin{equation}
 \frac{d p_j(t)}{d t} = -\alpha p_j
	+ \alpha \sum_{k} p_k(t) f_{j-k},
\end{equation}
with $f_k = f(k\Delta x) \Delta x$. A form of the Fokker-Planck equation
accurate up to the second order in space can be written with $f_k=0$ for all $k
\notin \{-1,0,1\}$ \footnote{It is also possible to write an equation where
$f_k$ is different of zero only when $k\in{0,1,2}$. However, this formulation
involve a negative $f_1$ which, besides having no probabilistic interpretation
leads to numerical instabilities.}. On the other hand, the IRRW implies in
$f_k\ge 0$ for $k\ge 0$ and $f_k=0$ for $k<0$. Similarly, it requires
\begin{equation}
 \begin{cases}
    f(\phi) \ge 0  & \text{for } \phi \ge 0 \\
    f(\phi) = 0  & \text{for } \phi<0
 \end{cases} .
\end{equation}
The above conditions exclude the Fokker-Planck equation or any form of
Eq. (\ref{FokkerPlanck}) involving $\phi$ derivatives.

By substituting Eq. (\ref{intEq}) in the time derivative of equations
(\ref{mu}) and (\ref{sig2}) we can describe the evolution of the
 average and variance of the p-age as
\begin{equation}
 \frac{d \mu(t)}{dt} = \alpha \langle x \rangle_f, \hspace{1cm}
 \frac{d \sigma^2(t)}{dt} = \alpha \langle x^2 \rangle_f
\end{equation}
where
\begin{equation}
\langle g(x) \rangle_f = \int g(x) f(x) \,dx.
\end{equation}
The comparison of these equations with Eq. (\ref{dmdt}) leads to the
conclusion that $\alpha$ and the function $f(x)$ must be such that
\begin{equation} \label{nuGamma1}
v = \alpha \langle x \rangle_f \hspace{1cm}
D = \alpha \langle x^2 \rangle_f .
\end{equation}
Whatever function $f(x)$ is, if the values of $\langle x \rangle_f$ and $\langle
x^2 \rangle_f$ are definite, it is always possible to rescale $f(x)$ and choose
$\alpha$ so that the above equations are satisfied.

If $f(x)$ doesn't include delta functions centered in 0 then $f(x)\neq 0$ for
values of $x\neq 0$ and it must, according to Eq. (\ref{nuGamma1}),
extends for at least a finite $L>0$. Consequently, the IRRW represented by
Eq. (\ref{intEq}) covers finite distances in an infinitesimal time
interval, meaning that the particles have infinite speed. Although this fact
may seem strange, it should not be a serious problem, since even the
well-accepted Fokker-Planck equation presents that property.

Similar use of an integral can be find in \cite{CunAl}, where the usual
nonlinear term of the Fisher equation was integrated over space to express
nonlocal competition. In our model, the integral is over the physiological age,
but it is still referred to as the nonlocal term.

The integral form (\ref{intEq}) is not suitable for numerical calculations,
since an integral must be evaluated at each discrete point $\phi$. Fortunately,
there exist functions for which the integral at $\phi$ may be quickly
calculated from the value at $\phi-\Delta\phi$. Two of these functions are the
linear and exponential functions.

\section{Speed states formulation}

Besides using a non-local formulation and a scalar field, the IRRW can be
described by a multi-component field $\mathbf{p}(\phi,t) = \{p_1(\phi,t),
\cdots, p_n(\phi,t)\}$ with the probability density given by
\begin{equation}
p(\phi,t) = \sum_i p_i(\phi,t) .
\end{equation}
Each component $p_i$ corresponds to a population that moves without dispersion
with speed $v_i$, and switch from state $i$ to state $j$ with rate $T_{ji}$.
They are subject to a local dynamic equation
\begin{equation} \label{multComp}
 \frac{d p_i(\phi,t)}{dt} = - v_i \frac{dp_i(\phi,t)}{d\phi}
	+ \sum_j T_{ij} p_j(\phi,t).
\end{equation}

If the above equation are integrated on $\phi$ disregarding the boundary
effects, the master equation is found to be
\begin{equation} \label{master}
 \frac{d p_i(t)}{dt} = \sum_j T_{ij} p_j(t),
\end{equation}
where
\begin{equation}
p_i(t) \equiv \int p_i(\phi,t) \, d\phi .
\end{equation}
Probability conservation imposes $\sum_i T_{ij} = 0$. {\vermelho By summing
Eq. (\ref{multComp}) over $i$ we arrive to the dynamics equation for the population density
\begin{equation} \label{dPptdt}
 \frac{\partial p(\phi,t)}{\partial t} =
	- \sum_i v_i \frac{\partial p_i(\phi,t)} {\partial \phi}.
\end{equation} }

Being  $\mathbf{T}$ a Markov matrix for continuous time, its eigenvalues are
all negative, except by one non-degenerated eigenvalue $\lambda^0 =0$. The
components $p_i^0$ of this stationary normalized eigenvector are all greater
than or equal to zero. If $\lambda^1$ is the second greatest eigenvalue,
$\mathbf{p}(t\gg 1/|\lambda^1|) \approx \mathbf{p^0}$.

Even after the steady state of Eq. (\ref{master}) is reached, the
convective term of Eq. (\ref{multComp}) constantly moves the local
population away from the equilibrium. After the transient is gone, the
stationary state $\mathbf{p}(t) = \mathbf{p^0}$ can be used to derive the
dynamics of
\begin{equation}
\mu_i(t) \equiv \frac{1}{p_i^0} \int \phi \; p_i(\phi,t)\,d\phi
\end{equation}
from Eq. (\ref{multComp}), resulting in
\begin{equation}
\begin{split}
\label{dmuidt}
 \frac{d\mu_i(t)}{dt} &= v_i + \frac{1}{p_i^0} \sum_j T_{ij} p_j^0 \mu_j(t)\\
&= v_i + \frac{1}{p_i^0} \sum_j T_{ij}^0 \mu_j(t).
\end{split}
\end{equation}
{\vermelho Since $\mathbf{T}$ is a continuous time Markov matrix, $T^0_{ij}
\equiv T_{ij} p^0_j$ share the same property. Furthermore, once $\mathbf{p^0}$
is an eigenvector of $\mathbf{T}$ with eigenvalue $0$, a vector with all
elements identical will be an eigenvector of $\mathbf{T^0}$ with the same
eigenvalue.}

In this stationary regime, the dynamics of
\begin{equation}
\mu(t) = \sum_i p_i^0 \mu_i(t)
\end{equation}
can be derived from Eq. (\ref{dmuidt}) or by integrating Eq. (\ref{dPptdt}):
\begin{equation} \label{dmutdt}
 \frac{d \mu(t)}{d t} = \sum_i v_i p_i^0  \equiv v^0 .
 \end{equation}
The solution of that equation is
\begin{equation}
\mu(t) = m^0 + v^0 t.
\end{equation}
The unceasing exchange of particles among the speed states at Eq.
(\ref{multComp}) connects the population of these states in bundles moving with
speed $v^0$.

{\vermelho  The motion of the center of mass of each state depends not only on
its own speed but also on the other states' centers of mass positions.
The motion is governed by the nonhomogeneous Eq. (\ref{dmuidt})
which possesses the particular solution
\begin{equation} \label{muit}
  \mu_i(t) = m_i + v^0 t ,
\end{equation}
with $m_i$ satisfying
\begin{equation}
 \frac{1}{p_i^0} \sum_j T_{ij}^0 m_j = v^0 - v_i. \label{mu0}
\end{equation}

The homogeneous part of Eq. (\ref{dmuidt}),
\begin{equation}
 \frac{d\mu_i(t)}{dt} = \frac{1}{p_i^0} \sum_j T_{ij}^0 \mu_j(t) , \label{homog}
\end{equation}
depends on the eigenvalues of $T_{ij}^0/p_i^0$. The eigenvalues of
$\mathbf{T^0}$ are all negative but the zero one. The same property is shared
by $T_{ij}^0/p_i^0$ because $p_i^0>0$. The homogenous solution disappears after
a while, except for the eigenvector $\mu_i = m^0$ of the null eigenvalue, which
can be included in $m_i$ of eq. (\ref{muit}). Therefore, eq. (\ref{muit}) is
the asymptotic solution of eq. (\ref{dmuidt}).}

Although the population average position at each speed state can be ahead or
behind the average position of the whole group, they all move forward with the
same average speed  $v^0$. The null eigenvalue corresponding to the constant
eigenvector of $\mathbf{T^0}$ means that the vector $m_j=\text{const}$ can be
added to the solution of (\ref{mu0}), reflecting the translational invariance
of the system.

After some algebra, the time evolution of $\sigma^2(t)$ can be obtained from
Eq. (\ref{multComp}) and (\ref{muit}) when $t\gg1/|\lambda^1|$. Thanks to
this result, together with Eq. (\ref{dmutdt}), expressions for $v$ and
$D$ are found
\begin{align} \label{NuGamma2}
 v &= \sum_i v_i p_i^0 ,\\
 D &= 2 \sum_i (m_i-m^0) v_i p_i^0.
\end{align}
The subtraction of $m^0$ from $m_i$ removes any dependence on a uniform
translation related to the zero eigenvalue of $\mathbf{T^0}$.

The simplest possible system able to accommodate
Eq. (\ref{NuGamma2}) is
\begin{equation}
 \mathbf{T} = \left[\begin{matrix} -k & k\\ k &-k \end{matrix}\right]
 \hspace{1cm}
 \mathbf{v} = \left[\begin{matrix} 0\\\nu \end{matrix}\right]. \label{Tv}
\end{equation}
The eigenvalues and eigenvectors of $\mathbf{T}$ are
\begin{align}
 \lambda^0 =0, \hspace{1cm}&
 \mathbf{p^0} = \frac{1}{2}\left[\begin{matrix}1 \\ 1 \end{matrix}\right], \\
 \lambda^1 = - 2k, \hspace{1cm}&
 \mathbf{p^1} = \frac{1}{2}\left[\begin{matrix}1 \\ -1 \end{matrix}\right].
\end{align}
When these values are substituted in Eq. (\ref{mu0}) the values of $m_i$
are obtained (the two equalities of the linear system are not linearly
independent)
\begin{equation}
 \mathbf{m} = \left[\begin{matrix} m^0 - \nu/4k \\ m^0 + \nu/4k
	\end{matrix}\right].
\end{equation}

The average value of $\phi$ of the whole population in $t=0$ would be $m^0$,
provided that equilibrium was already reached at that time. The
average position of the zero speed individuals is behind the average
population's position, while the
$\nu$ speed individuals move ahead of the group.

Concluding, the constants for the two speed states system are
\begin{equation}
 v = \frac{\nu}{2} \hspace{1cm} D = \frac{\nu^2}{4k} .
\end{equation}

\section{Comparing the models}

{\vermelho The time evolution of $\mu(t)$ and $\sigma^2(t)$ obtained by numeric
integration of the three models can be seen in figures \ref{fig:estatist}$a$
and \ref{fig:estatist}$b$. When $t\lesssim 1$, $\sigma^2(t)$ of the two speeds
model slightly deviates from the linear behavior, which may be due to the
transients of the homogeneous solution of Eq. (\ref{dmuidt}). Except for
that, these two curves perfectly agree with the analytic results.

\begin{figure}
 \centering
 \includegraphics{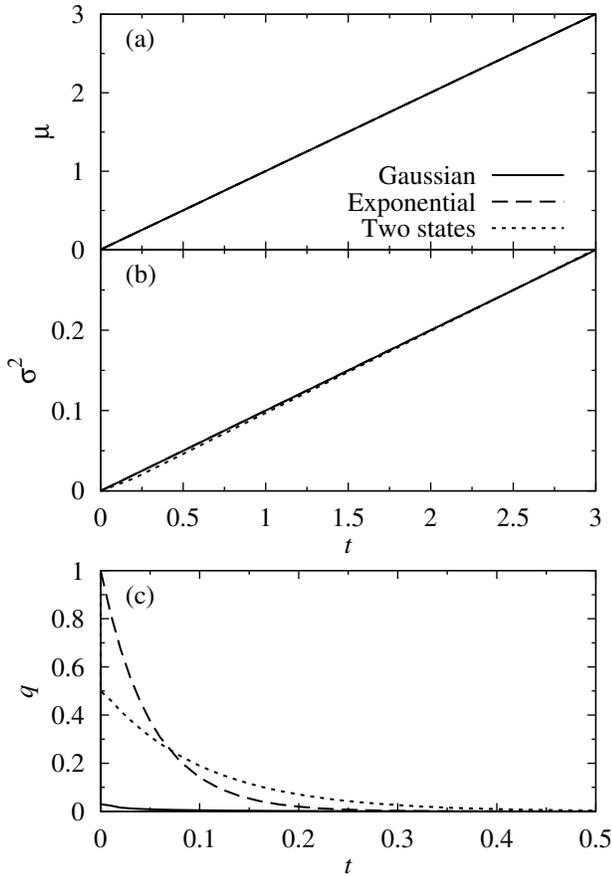}
 \caption{Results of the numerical integration according to the
Fokker-Plank equation (Gaussian), non-local IRRW with exponential
function and two speed states model, with $v=1$ and $D=0.1$.  The initial distribution was
$p(\phi,0) = \delta(\phi)$, with the population of the two speed model already
in stead state: $p_0(t=0)=p_1(t=0)$. a) Population average position. b) Population position
variance. c) Value of $q(t)$ as defined in Eq. (\ref{q}). } \label{fig:estatist}
\end{figure} }

Figure \ref{fig:3form}$a$ shows that the three approaches result in very
different population distribution for short time scales. {\vermelho In the
Fokker-Plank dynamics the initial delta distribution becomes immediately
Gaussian, while the disappearing of the delta takes some time in the other two
models. Aiming to a quantitative evaluation of this phenomenon, we define the
quantity
\begin{equation} \label{q}
q(t) = \int \delta(\phi)\: p(\phi,t)\, d\phi .
\end{equation}
The evolution of $q(t)$ can be seen in figure \ref{fig:estatist}$c$.

\begin{figure}
 \centering
 \includegraphics{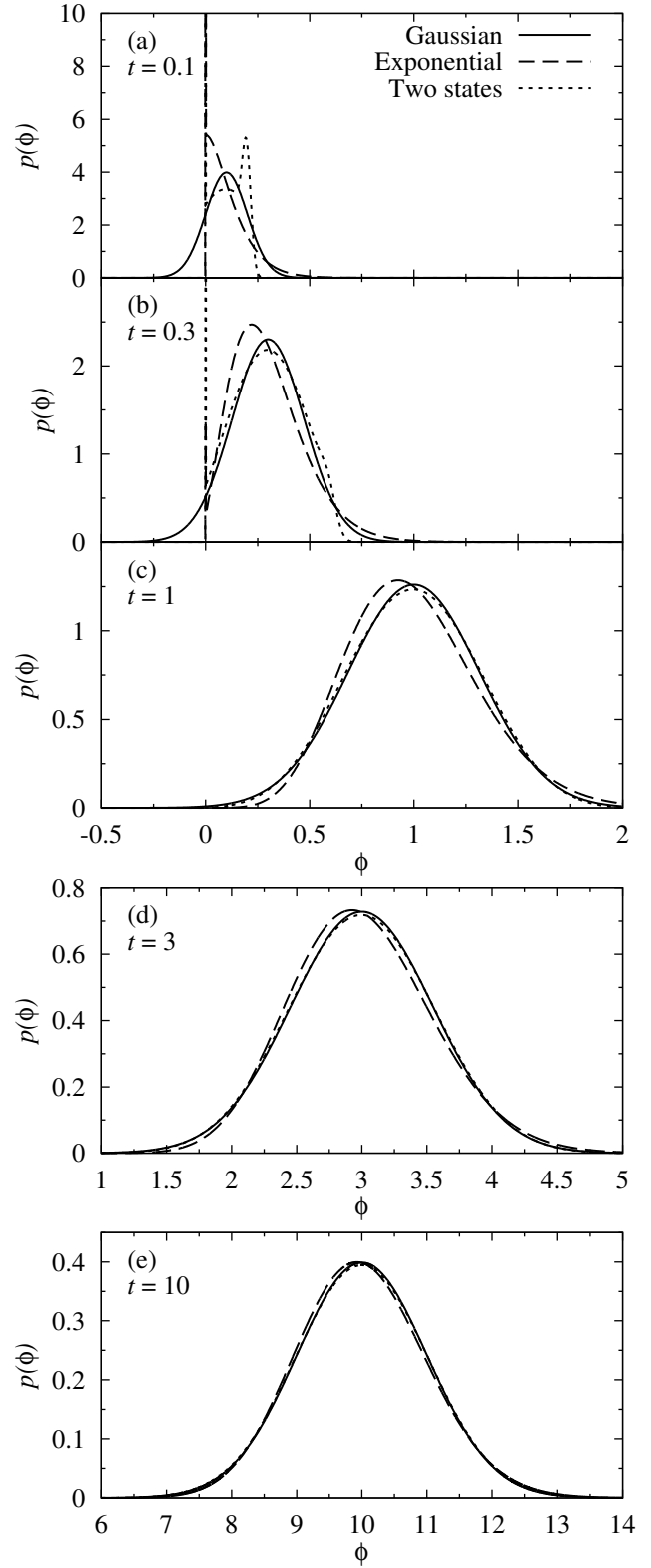}
\caption{Population distribution resulting from the
same numerical integration of figure \ref{fig:estatist} at different
instants for the three models.} \label{fig:3form}
\end{figure}

The instantaneous disappearing of the delta function in the Fokker-Plank
equation, resulting in $q(0)\approx0$, is only possible due to the singularity
of the laplacian of the delta function. The nonexistence of a similar
singularity in the non-local model prevent the same effect in this model. In
the two speeds model, the singular derivative of $dp/d\phi$ in the $v>0$ state
instantly moves the population at this state away from $\phi=0$, resulting in
$q(0)=1/2$.

If we apply the frame of reference change $x \rightarrow x-\nu t$ in the two
states model, the only effect would be replacing $\mathbf{v}$ in eq. (\ref{Tv})
by
\begin{equation} \label{v1}
 \mathbf{v} = \left[\begin{matrix} -\nu \\0 \end{matrix}\right].
\end{equation}
The resulting distribution with such $\mathbf{v}$ would be translations of the
figure \ref{fig:3form} distributions. On the other hand, symmetry analysis
implies that the distributions resulting from eq. (\ref{Tv}) or from eq.
(\ref{v1}) should be symmetric by reflection, and a delta should be present at
the right of the bells seen in figures \ref{fig:3form}$a$ and
\ref{fig:3form}$b$. A trace of such delta function is present on figure
\ref{fig:3form}$a$, but complectly disappeared on figure \ref{fig:3form}$b$,
due to the numerical discretization.

The asymmetry of the distribution resulting from the non-local model is visible
when we compare that distribution, in figure \ref{fig:3form}, with the perfectly
symmetric the Fokker-Planck distribution. That asymmetry is not
a surprise, since the non-local Eq. (\ref{intEq}) is not symmetric.

After some time, the distribution at $\phi>0$ of the IRRW models look like a
truncated bell, but the curves become bell shaped only after the distribution
moves far enough from the initial position. The two states model becomes
Gaussian faster than the non-local model. This happens despite the initial
asymmetry generated in the two states model by the numerical error. }

Regarding numerical implementation, the Fokker-Planck equation usually involves
inverting a tri-diagonal matrix when using second-order accuracy on time. The
two speeds model methods doesn't, since it can be implemented using upwind
differencing.

When applying the boundary conditions to the Fokker-Planck equation, it is
necessary to prevent the movement in the wrong direction. One doesn't need to
worry about such matters when using the IRRW models. On the other hand, it is
awkward to include far-reaching effects of the non-local formulation in the
boundary conditions, mainly between development stages.

Except for involving more than one field, the numerical implementation of the
two speed states model is overall more convenient than the Fokker-Planck and
the non-local models.

Only specially planned experiments, or a deeper understanding of the aging
process, will decide which, if any, of the three models discussed in this work
is correct. Whatever the answer is, it won't invalidate the use of the other
methods as a convenient approximation, not to mention the fact that many
different behaviors can be accommodated in the freedom of choosing $f(x)$ in
Eq. (\ref{intEq}) or $v_i$ and $T_{ij}$ in Eq. (\ref{multComp}).

\section{conclusion}

{\vermelho Two basic equations describing the evolution of the population
distribution were proposed. As already mentioned, the main problem with the
Fokker-Planck equation is its non-physical properties, namely, negative
variation of $\phi$ and the infinite speed. The non-local formulation solves
the first problem but not the second, while the speed states formulation solves
both of them.

An advantage of the non-local and the two speed models over the Fokker-Plank
equation is their convenience for certain numerical implementations. The
analytical expressions for the mean and the variance of the p-age help to
understand the role played by the parameters of the dynamic equations.

Several other biological phenomena could be included in a more detailed
formulation. Some of them, such as the  age structure, the quiescence, and the
dependence of the biological development on the temperature, could be included
by making the equation parameters dependent on temperature and age. Other
phenomena such as death, reproduction,  and diapause, require the introduction of
new terms in the dynamic equation. Introducing more complex effects such as
spatiality and inter species interaction can only be done by defining new
independent variables and fields.}

Real situations demand taking into account some of these phenomena, requiring several additional
assumptions and the determination of the corresponding parameters. Possible
usages include plague control, epidemics, ecological management, demography,
etc. They go well beyond the scope of this paper, which presents the equations
ruling the intra-stage development.

Most models of genetic agents explored by physicists use c-age, either discrete
\cite{Pen} or continuous \cite{HwaKraRed}. These models could be extended by
incorporating the p-age, particularly, the IRRW presented here.

\begin{acknowledgments}
I am grateful to Fernando A. Oliveira for the thoughtful discussions and to the
National Council for Scientific and Technologic Development - CNPq, for the
financial support.
\end{acknowledgments}


%
\end{document}